\documentstyle[12pt]{article}
\begin{document}
\title{TOPOLOGY AND QUANTIZATION
\footnote
{The paper is partially supported by KBN, Grant No 2P 302 087 06,\\
and by the Katholischer Akademischer Auslander Dienst, Bonn, Germany}}
\author{W{\l}adys{\l}aw Marcinek
\\ Institute of Theoretical Physics, University
of Wroc{\l}aw,\\ Pl. Maxa Borna 9, 50-204  Wroc{\l}aw,\\
Poland}
\date{}
\maketitle
\begin{abstract}
\noindent
A simple algebraic model of charged particle coupled to singular 
magnetic field is given. Quantization is described as gradation
by certain abelian group $G$. Statistics is determined by a commutation 
factor $\lambda$ on the grading group $G$. Composite fermions and composite 
bosons are described in an unified way as $Z_2\oplus...\oplus Z_2$-graded 
$\lambda$-commutative algebras. 
\end{abstract}
\def\ot{\otimes}
\def\ra{\longrightarrow}
\def\ca{{\cal A}}
\def\cl{{\cal L}}
\def\cw{{\cal W}}
\def\cc{{\cal C}}
\def\cb{{\cal B}}
\def\cm{{\cal M}}
\def\pr{(.|.)}
\def\sc{<.|.>}
\def\sh{S_{\chi}}
\def\ps#1,#2,{\Psi_{{#1}{,}{#2}}}
\def\a{\alpha}
\def\b{\beta}
\def\la{\lambda}
\def\g{\gamma}
\def\ep{\epsilon}
\def\si{\sigma}
\def\vp{\varphi}
\def\th{\Theta}
\def\sa{S_{\ca,\ca}}
\def\1n{^{(1)}}
\def\2n{^{(2)}}
\def\3n{^{(3)}}
\def\p{\partial}
\def\ba{\begin{array}}
\def\ea{\end{array}}  
\def\be{\begin{equation}}
\def\ee{\end{equation}}
\newpage
% section 1
\section{Introduction}
It is well-known that there is a mathematical formalism for the
description of particle systems in a low dimensional space $\cm$ 
based on the notion of the braid group $B_n (\cm)$, see Ref.\cite{Wu,I} 
for example. In this this approach the configuration space for the system 
of $n$--identical particles moving on a manifold $\cm$ is 
$Q_{n}= {\left(\cm^{\times n} - D\right) }/{S_{n}}$,
where $D$ is the subcomplex of the Cartesian product $\cm^{\times n}$ on 
which two or more particles occupy the same position and $S_{n}$ is the 
symmetric group. The group $\pi_{1}\left( Q_{n}(M)\right)
\equiv B_{n} (M)$ is known as the $n$--string braid group on $\cm$.
The statistics of the given system of particles is described by the
group $\Sigma_n (\cm)$ which is a subgroup of the braid group $B_{n}$
corresponding to interchanges of two particles, one by another.
Quantizations are described by unitary representations of the braid group. 
This picture breaks upin the one-dimensional case. The difficulty arises 
with the proper definition of the group $\Sigma_n (\cm)$, see \cite{jsw}.
In this paper we going to present a proposal for quantum 
model of charged particle in low-dimensional space 
$\cm$ with perpendicular singular magnetic field. Our model
is pure algebraic. It is based on the notion of the homotopy
theory and $G$-graded structures.
Our fundamental assumption is that every 
charged particle such as electron coupled 
to the singular magnetic field is transformed into a composite
system which consists a charge and certain number of magnetic fluxes.
Every magnetic flux can be compensated by the particle such that
the effective magnetic field is zero or can not be compensated.
In the first case we say that we have a quasiparticle state,
in the second - a quasihole one.
We also assume that there is in average $N$ fluxes per particle.
This means that the filling factor is $v = \frac{1}{N}$.
If the number $N$ of fluxes is even, then the system can be identified 
as composite fermions. If $N$ is odd, then we obtain composite bosons.
The paper is organized as follows. In Sect. 2 we study the quantizations
of our particle moving in the singular magnetic field.
Note that the "effective" configuration space for our particle is 
$\cm \equiv S^1 \times \ldots \times S^1$. The fundamental group
of $\cm$ is denoted by $\pi_1 (\cm, m_0)$, the base point $m_0$ is the
initial point of our particle in the moment $t =  0$. Let $G$ be certain
subgroup of the fundamental group, then there is a group algebra 
$H := I\!\!\!\! {\bf C}G$ which has a Hopf algebra structure.
In our algebraic approach all possible quantizations are described 
by $H$-comodule coaction on a given Hilbert space $E$. It is known
that $H$-coaction on $E$ is equivalent to the $G$-gradation of $E$,
see \cite{mon}. Hence quantization can be also described as certain 
gradation of $E$. The statistics is determined by the normalized 
bicharacter (a commutation factor) $\la$ on the grading group $G$, 
i.e. a mapping $\la : G \times G \ra I\!\!\!\! {\bf C} \setminus \{0\}$ 
satisfing some condition given in Sec. 2. 
The algebra of quantum states for our composite system of particle
and magnetic fluxes is described in general case as $G$-graded 
$\la$-commutative algebra in Sec. 3, where $G$ is an arbitrary
abelian group and $\la$ is a commutation factor on it. 
In Sec. 4 the algebra of states for composite fermions or composite 
bosons is described in more details as $G$-graded $\la$-commutative 
algebra $\ca_{\la}$ which is generated by $x_i$, $(i = 1,...,N)$ such 
that we have the following commutation relations
\be
\ba{c}
x_{i} \ x_{j} = \la_{ij} \ x_{j} \ x_{i},
\label{lco}
\ea
\ee
where the factor $\la = \la_{ij}$ is defined by the following formula
\be
\ba{c}
\la_{ij} = -(-1)^{\delta_{ij}} \ q^{\Omega_{ij}}
\label{acom}
\ea
\ee
for $i, j = 1, 2,...,N$, $\Omega_{ij} = 1$ for $i < j$,
$\Omega_{ij} = - \Omega_{ji}$, $q = exp(\pi i N)$ and the
grading group is $G \equiv Z_2 \oplus...\oplus Z_2$ (N-sumands).
Such algebra is realized as certain subalgebra in the tensor
product of paragrassmann (or parabose) algebra by Clifford
algebra. Some physical consequences of our model are given.
Note that similar graded structures has been studied previously
by several authors but in different context, see Ref. 
\cite{luk,rit,ohn} for example.
The paper is continuation of the author's previous ones 
\cite{WM4,WM5,WM6,WM8}. 
% S E C T I O N 2
\section{Quantization as gradation}
Let us consider an arbitrary charged particle moving on the space $\cm$.
We assume that $\cm$ is path-connected topological with base point $m_0$.
All possible classical trajectories of the particle are path in $\cm$.
We denote by $P_m\cm$ the space of all homotopy classes of paths which
starts at $m_0$ and and end at arbitrary point $m \in \cm$. It is known
that the union $\cup_m P_m\cm = P\cm$ is a covering space 
$P = (P\cm, \pi, \cm)$. The projection $\pi : P\cm \ra \cm$
is given by
\be
\ba{ccc}
\pi (\xi) = m &\mbox{iff}&\; \xi \in P_m \cm .
\ea
\ee
The homotopy class $P_{m_0}\cm$ of all paths which start at $m_0$
and end at the same point $m_0$, (i.e. a loop space) can be
naturally endowed with a group structure. This group is known
as the fundamental group of $\cm$ at $m_0$ and is denoted by
$\pi_1 (\cm, m_0)$. Generators of the fundamental group are
denoted by $\si_i$.
In our case we have $\cm = S^1 \times \ldots \times S^1$ and
the fundamental group is
\be
\pi_1 (\cm, m_0) = Z \oplus...\oplus Z \;\;\; \mbox{(N-sumands)}.
\ee
Let $G$ be a subgroup of the fundamental group $\pi_1 (\cm, m_0)$.
We assume that quantum states of magnetic field can be represented
as linear combinations of elements (over a field $I\!\!\!\! {\bf C}$
of complex numbers) of the group $G$. It is known such that linear 
combinations of elements of certain group $G$ form a group algebra
$H := I\!\!\!\! {\bf C}G$. It is also known that there is a Hopf 
algebra structure defined on the group algebra $H$.
Let us denote by $E$ a Hilbert space of quantum states of particle
which is not coupled to magnetic field. Quantum states of particle
coupled to our singular magnetic field is described by the tensor
product $E \ot H$. Every attaching of magnetic fluxes to the particle 
moving in singular magnetic field can be represented by certain coaction
$\rho_E$ of the Hopf algebra $H$ on the space $E$, i.e. by a linear
mapping
\be
\rho_E : E \ra E \ot H,
\ee
which define a (right-) $H$-comodule structure on $E$. 
Note that if $E$ is a $H$-comodule, where $H = I\!\!\!\! {\bf C}G$,
then $E$ is also a $G$-graded vector space, i.e
\be
E = \bigoplus\limits_{\a \in G} \ E_{\a}.
\ee
The family of all $H$-comodules forms a category $\cc = \cm^H$.
The category $\cc$ is braided monoidal.
The monoidal operation in $\cc = \cm^H$ is given as the following
tensor product of $H$-comodules
\be
\rho_{E \ot E} = (id \ot m) \circ (id \ot \tau \ot id)
\circ (\rho_E \ot \rho_E),
\ee
where $\tau : H \ot E \ra E \ot H$ is the twist, 
$m : H \ot H \ra H$ is the multiplication in $H$.
The braid symmetry 
$\Psi \equiv \{\ps U, V, : U \ot V \ra V \ot U; U, V \in  Ob \cc\}$
in $\cc$ is defined by 
\be
\ps U, V, (u \ot v) = \la (\a, \b) \ v \ot u
\ee
for $u \in U_{\a}$, $v \in V_{\b}$, $\la$ is a bicharacter on the
group $G$, i.e. a mapping 
$\la : G \times G \ra I\!\!\!\! {\bf C} \setminus \{0\}$
such that
\be
\ba{c}
\la (\a + \b, \g) = \la (\a, \g) \ \la (\b, \g),
\la (\a, \b + \g) = \la (\a, \b) \ \la (\a, \g).
\ea
\ee
for all $\a, \b, \g \in G$. In this paper we restrict our attention
to abelian grading group $G$ and normalized bicharacters (called by
Scheunert \cite{sch} a commutation factors on $G$) such that
\be
\la (\a, \b) \ \la (\b, \a) = 1.
\ee
In this particular case the category $\cc$ becomes symmetric.
% S E C T I O N 3
\section{Algebra of states}
It is interesting that there exist an algebra $\ca$ in the symmetric
monoidal category $\cc$ which is $G$-graded
\be
\ca = \bigoplus\limits_{\a \in G} \ \ca_{\a} ,\;\;\;
\ca_{\a} \ \ca_{\b} \subset \ca_{\a + \b},
\ee
and $\la$-commutative
\be
x_{\a} \ x_{\b} = \la (\a, \b) \ x_{\b} \ x_{\a}
\ee
for homogeneous $x_{\a} \in \ca_{\a}$, $x_{\b} \in \ca_{\b}$.
Note that the algebra $\ca$ is also a coalgebra
\be
\Delta x_{\g} = \sum\limits_{\a + \b = \g} \
\sum\limits_i \ x_{\b ,i} \ x_{\a ,i}.
\ee
Moreover, one can see that there is a graded Hopf algebra structure 
on $\ca$, see \cite{mon}. We use here the so-called standard gradation 
for simplicity. In this gradation the algebra $\ca$ is generated by 
$x_i$, $(i = 1,...,N)$ such that
\be
\ba{c}
x_{i} \ x_{j} = \la_{ij} \ x_{j} \ x_{i},
\label{laco}
\ea
\ee
where the factor $\la_{ij}$ is defined by the following formula
\be
\ba{cc}
\la_{ij} := \la (\si_i, \si_j),&\;i,j = 0, 1,...,N,
\label{wel}
\ea
\ee 
$\si_i$ for $(i = 1,...,N)$ are generators of $G$, 
$grade(x_i) = \si_i$, and it is natural to assume that 
$\si_0 \equiv e$, $e$ is the neutral element in $G$. It is obvious that 
$\la_{ij} \in I\!\!\!\! {\bf C}\setminus\{0\}$ for every $i,j = 1,...,N$
and
\be
\ba{c}
\la_{ii} = \pm 1,\;\;\;  \la_{ij} \ \la_{ji} = 1,
\;\;\; \la_{0j} = \la_{i0} = \la_{00} = 1.
\label{sar}
\ea
\ee
The set of numbers $\{\la_{ij} : i,j = 0, 1,...,N\}$ is said to be
{\it a relative sign} or {\it relative phase}.
Every element $x_{\a}$ of the algebra $\ca$ can be given in the
following form
\be
\ba{c}
x_{\a} = x_1^{\a_1},..., x_N^{\a_N}
\label{stat}
\ea
\ee
for $\a = \Sigma_{i=1}^N \ a_i \si_i$, $x_0 \equiv {\bf 1}$, where
${\bf 1}$ is the unit in $\ca$. In physical interpretation
every element $x_{\a}$ of $\ca$ describe certain configuration
of charged particle $x$ coupled to our singular magnetic field.
Generators $x_i$ of $\ca$ correspond to particle coupled to
single magnetic flux at the point $s_i$ in the Landau lowest lewel. 
The unit ${\bf 1}$ of the algebra $\ca$ describe the particle which 
is not coupled to the magnetic field. The monomial $x_i x_j$
describe a particle coupled to two magnetic fluxes at $s_i$ and
$s_j$ simultaneously. It is obvious that $(x_i)^2$ corresponds
to particle coupled to two fluxes at the same point $s_i$.
A charged particle equipped with magnetic flux is said to be
{\it a quasiparticle}. The particle coupled to two magnetic
fluxes at two different point is understood as a system of two
different quasiparticles. It follows from our above interpretation 
that the monomial $x_i x_j$ describe particle coupled to two magnetic 
fluxes, i. e. a system of two different quasiparticles $x_i$ and $x_j$. 
We assume that quasiparticles are identical. If we exchange these 
quasiparticles one by another, then we obtain the system $x_j x_i$
which must be equivalent to $x_i x_j$.  The only difference can be 
in phase. In fact the $\la$-commutativity (\ref{laco}) means that 
really the only difference is in the phase $\la_{ij}$. This also 
means that our quasiparticles have his own statistics. 
Let $x_{\a}$ be an arbitrary element of the algebra $\ca$ of the 
form (\ref{stat}). If $\a_i = 0$ for certain $i$, then $x_i^0 = {\bf 1}$,
and we say that we have a quasihole at $s_i$. In this way an element 
$x_{\a} \in \ca$ describe a system of quaisparticles and quasiholes.

It is known that every commutation factor $\la$ on an arbitrary
grading group $G$ can be given in the following form
\be
\ba{c}
\la (\a, \b ) = (-1)^{(\a |\b )} \ q^{<\a |\b >},
\label{dom}
\ea
\ee
where $(-|-)$ is an integer-valued symmetric bi-form on $G$,
and $<-|->$ is a skew-symmetric integer-valued  bi-form on $G$,
$q$ is some complex parameter \cite{zoz}. Note that $q$ in general 
is not a root of unity, but in the particular case when 
$G = Z_m \oplus...\oplus Z_m$
$q$ must be the $m$-th root of unity.
%            S  E  C  T  I  O  N      4
\section{Commutation factor for composite bosons and fermions}
Let us consider the grading group and commutation factor corresponding
to our composite system of particle and magnetic fluxes.
We use here the so-called standart gradation of Ref. \cite{WM4}. 
In this gradation the grading group $G$ is in general equal to 
$Z^N := Z \oplus...\oplus Z$ ($N$-sumands), $N = 0, 1,...$; 
$Z^0 \equiv \{e\}$ is the trivial group, and $Z^1 \equiv Z$ is the 
group of all integers. In physical interpretation $N$ denote the 
number of singular points. Let us calculate explicite the relative 
phase $\la_{ij}$ for our model. It follows immediately from formule
(\ref{wel}) and (\ref{dom}) that we have the following general
relation
$$
\ba{c}
\la_{ij} = \la (\si_i, \si_j) =  (-1)^{A_{ij}} \ q^{\Omega_{ij}},
\label{pha}
\ea
$$
where $A_{ij} := (\si_i|\si_j)$, and $\Omega_{ij} := <\si_i|\si_j>$
are integer-valued matrices such that $A_{ij} = A_{ji}$ and
$\Omega_{ij} = - \Omega_{ji}$. The above relative phase contains
two factors: $\la_{ij}^{st} := (-1)^{A_{ij}}$ and 
$\la_{ij}^{A-B} := q^{\Omega_{ij}}$. In our physical interpretation
$\la_{ij}$ describes the phase corresponding to interchanging of two
qusiparticles $x_i$ and $x_j$. Remember that quasiparticles $x_i$ and 
$x_j$ are in fact the same particle $x$ but equipped with two fluxes.
It is natural to assume that the first factor describes the statistics 
of the original particle $x$ (i.e. the particle under consideration, 
without magnetic fluxes) and the second one - the Aharamov-Bohm phase 
for two interchanging quasiparticles. If our particle is electron, 
then we must assume that 
$\la_{ii}^{st} \equiv -1$ and $\la_{ij}^{st} \equiv 1$ for all $i \neq j$,
This means that $A_{ij} = \delta_{ij}$ and we have
$\la_{ij}^{st} := (-1)^{\delta_{ij}}$. For $\la_{ij}^{A-B}$ we assume that
\be
\la_{ij}^{A-B} := \exp
\left[
\ba{c}
\pi i (\frac{e}{h} \Phi + 1)
\ea
\right],
\label{kuk}
\ee
for $i < j$, $e \neq 0$ is the electric charge, $h$ is the Planck constant.
For $i > j$ we have $\la_{ij}^{A-B} = (\la_{ji}^{A-B})^{-1}$, 
$\la_{ii}^{A-B} = 1$.
It follows immediately from the above formula (\ref{kuk}) that for 
\be
\ba{c}
\Phi = N \Phi_0 = N \frac{h}{e}, 
\label{phi}
\ea
\ee
where $N = 0, 1, 2,...$ we have
$$
\la_{ij}^{A-B} = \exp
\left[
\ba{c}
\pi i (N + 1) 
\ea
\right] =
\left\{
\ba{c}
- 1 \;\mbox{for}\; N \;\mbox{even}\\
+ 1 \;\mbox{for}\; N \;\mbox{odd}
\ea
\right..
$$
This means that we obtain $\la_{ij}^{A-B} = -(-1)^N$.
Hence the relative phase $\la_{ij}$ can be given in the form
\be
\ba{c}
\la_{ij} = -(-1)^{\delta_{ij}} q^{\Omega_{ij}},
\label{com}
\ea
\ee
where $q := \exp(\pi i N)$, $\Omega_{ij} = -\Omega_{ji} = 1$,
for $i \neq j$, $\Omega_{ii} = 0$, $i, j = 1, 2,...,N$. 
This means that we have $\la_{ij} = \pm 1$ and the grading 
group $G = Z^N$ can be reduced to the following group
\be
\ba{c}
G \equiv Z_2^N := Z_2 \oplus ... \oplus Z_2\;\;\;(N sumands).
\label{gru}
\ea
\ee
This also means that the charged particle can be coupled to one 
elementary magnetic flux $\Psi_e$ at every singular point $s_i$
and the filling factor is $v = \frac{1}{N}$.

Let $\ca \equiv \ca_{\la}$ be a $G$-graded $\la$-commutative 
algebra, where $G$ is given by the formula (\ref{gru}) and $\la$
by (\ref{com}). Here an arbitrary element $x_{\a}$ of $\ca_{\la}$
can be given in the form (\ref{stat}), where $\a_i = 0$ or $1$
for $i = 1,...,N$. If $\a_i = 0$, then we have quasihole, if
$\a_i = 1$, then we have quasiparticle. The number of quasiparticles
is: $m = \Sigma_{i=1}^{N} \a_i$.
Now we looking for the realization of the algebra $\ca_{\la}$
in the tensor product $\Lambda_s \ot C_{N}$ of two others algebras 
$\Lambda_s$ and $C_{N}$. We give the following ansatz
\be
x_i := \th_i \ot e_i
\ee
for $i = 1,...,N$. We assume that the algebra $\Lambda_s$ is generated 
by $\th_1,...,\th_N$ such that we have the following commutation relations
\be
\th_i \ \th_j = - \la_{ij} \ \th_j \ \th_i .
\ee
For $i=j$ we obtain that $\th_i^2=0$.
The algebra $C_{N}$ is the Cliford algebra. It is generated by 
$e_1,...,e_N$ such that we hahe the well-known relations
\be
\ba{cccc}
e_i \ e_j = - e_j \ e_i&\;\mbox{for}\;&i \neq j ,&\;
e_i^2 = 1
\ea
\ee
For the multiplication in the algebra $\ca \equiv \ca_{\la}$
we have
\be
x_{i} \ x_{j} = \th_{i} \ \th_{j} \ot e_i e_j .
\ee
Now let us study the above realization of $\ca \equiv \ca_{\la}$
in more details. Observe that for even $N$ we obtain 
\be
\ba{cc}
\th_i \ \th_j = \th_j \ \th_i,&\;\th_i^2 = 0.
\ea
\ee
It is interesting that such algebra can be represented by 
one grassmann variable $\th$, $\th^2 = 0$ as follows
\be
\ba{rcl}
\th_i = (0,...,&\th&,...,0).\\
&\uparrow&\\
&&\mbox{the i-th place}
\ea
\ee
We have
\be
\ba{rcccl}
\th_i \th_j = (0,...,&\th&,...,&\th&,...,0) = \th_j \th_i.\\
&\uparrow&&\uparrow&\\
&&\mbox{the i-th place}&&\mbox{the j-th place}
\ea
\ee
and
\be
\ba{rcl}
\th_i^2 = (0,...,&\th^2&,...,0) = 0.\\
&\uparrow&\\
&&\mbox{the i-th place}
\ea
\ee
For the quantum state $x_i x_j (i \neq j)$ representing particle
coupled to magnetic fluxes at two different points we have
\be
\ba{c}
x_i \ x_j = (0,...,\th,...,\th,...,0) \ot e_i \ e_j .
\ea
\ee
Now it is easy to see that the algebra $\ca_{\la}$ for even $N$ 
describes composite fermions. Let us consider the case of $N = 2$
in more details. In this case we have the following atates
\be
\ba{cc}
x_1 = (\th, 0) \ot e_1 ,&\;  x_2 = (0, \th) \ot e_2,
\label{sqa}
\ea
\ee
and
\be
\ba{c}
x_1 \ x_2 = (\th, \th) \ot e_1 \ e_2 .
\label{haf}
\ea
\ee
The filling factor for all these states (\ref{sqa}) and (\ref{haf})
is $v = \frac{1}{2}$. The states (\ref{sqa}) contain quasiholes.  
Note that the state (\ref{haf}) does not contain quasiholes 
and hence it is unique state corresponding for which the
magnetic field is completely compensated!

For odd $N$ we have
\be
\th_i \ \th_j = - \th_i \ \th_j \;\;\;\mbox{for}\; i \neq j .
\ee
For $i = j$ we obtain the identity $\th_i \th_i = \th_i \th_i$.
In this case the algebra $\ca_{\la}$ can also be represented by 
the variable $\th$ such that we have
\be
\ba{rcl}
\th_i = (0,...,&\th&,...,0).\\
&\uparrow&\\
&&\mbox{the i-th place}
\ea
\ee
We have
\be
\ba{rcccl}
\th_i \th_j = (0,...,&\th&,...,&\th&,...,0) = - \th_j \th_i.\\
&\uparrow&&\uparrow&\\
&&\mbox{the i-th place}&&\mbox{the j-th place}
\ea
\ee
This means that
\be
\th_i \ \th_j = 0
\ee
for all $i \neq j$. Observe that the quantum state $x_i x_j$ $(i \neq j)$
corresponding to particle coupled to magnetic fluxes at two different
points disappear
\be
x_i \ x_j = \th_i \ \th_j \ot e_i \ e_j = 0,
\ee
and the state describing the particle coupled to a few fluxes at
the same point is also impossible. In fact we have
\be
x_i^2 = (0,...,\th^2,...,0) \ot e_i^2 = 0.
\ee
This means that the state $x_i^2$ is not equipped with a flux.
Let us consider as an example the case of $N = 3$ i.e. the filling
factor is $v = \frac{1}{3}$. In this case we have the states
\be
\ba{ccc}
x_1 = (\th, 0, 0) \ot e_1 ,&\;  x_2 = (0, \th, 0) \ot e_2,&
\; x_3 = (0, 0, \th) \ot e_3 
\label{star}
\ea
\ee
which contain two quasiholes.
Observe that the following states
\be
\ba{ccc}
x_1 x_2 ,&\;  x_1 x_3,&\; x_2 x_3 .
\label{atar}
\ea
\ee
which contain one quasihole and the state
\be
\ba{c}
x_1 x_2 x_3
\label{ful}
\ea
\ee
which not contain quasiholes are impossiible. Hence in 
this case the single quasiparticle
states with two quasiholes are possible! This also means
that the Landau lowest lewel is $\frac{1}{3}$-filled.
In this way we obtain simple algebraic description of quantum 
states for particle in singular magnetic field which agree
with known facts. 

\vspace{3cm}

\noindent {\bf Acknowledgments}
The author would like to thank to the organizer of SSPCM96
for his kind invitation to Zajaczkowo, to K. Wieczorek
for discussions, and to A. Borowiec, for any others help.


\vspace{3cm}

\begin{thebibliography}{99}
\bibitem{Wu}  Y.S. Wu,J.Math.Phys. {\bf 52}, 2103, 1984
\bibitem{I} T.D. Imbo and J. March--Russel,
Phys. Lett.{\bf B252}, 84, 1990
\bibitem{jsw} L. Jacak, P. Sitko, K. Wieczorek, Anyons and composite
fermions, Scientific Papers of the Institute of Phsics of the Technical
University of Wroclaw No. 31, Monographs No. 20, Wroclaw 1995.
\bibitem{mon} S. Montgomery, Hopf algebras and their actions on rings,
Regional Conference series in Mathematics, No 82, AMS 1993.
\bibitem{wil} F. Wilczek, {\em Phys. Rev. Lett.} {\bf 48}, 114 (1982).
\bibitem{luk} J. Lukierski, V. Rittenberg, {\em Phys. Rev. Lett.} 
{\bf D18}, 385, (1978).
\bibitem{rit} V. Rittenberg, D. Wyler, {\em J. Math. Phys} {\bf 19},
 2193, (19780).
\bibitem{ohn} Y. Ohnuki, S. Kamefuchi, {\em J. Math. Phys.} 
{\bf 21}, 601 (1980).
\bibitem{anm} S. Majid, Anyonic quantum groups, in Spinors, 
Twistors and Clifford algebras, ed. by Z. Oziewicz et al, 
Kluwer Acad. Publ. (1993), p. 327.
\bibitem{bnm} S. Majid, {\em Czech. J. Phys.} {\bf 44}, 1073 (1994).
\bibitem{sch} M. Scheunert, {\em J. Math. Phys.} {\bf 20}, 712, (1979).
\bibitem{WM4} W. Marcinek, {\em Rep. Math. Phys.} {\bf 34}, 325, (1994).
\bibitem{WM5} W. Marcinek, {\em Int. J. Mod. Phys.} {\bf A10}, 1465 (1995).
\bibitem{WM6} W. Marcinek, {\em Rep. Math. Phys.} {\em 33}, 117, (1993).
\bibitem{WM8} W. Marcinek, {\em J. Math. Phys.} {\bf 35}, 2633, (1994).
\bibitem{zoz} Z. Oziewicz, Lie algebras for arbitrary grading group, in
Differential Geometry and Its Applications ed. by J. Janyska and D. Krupka,
World Scientific, Singapore 1990.
\end{thebibliography}
\end{document}